\documentclass{article}

\usepackage{microtype}
\usepackage{graphicx}
\usepackage{subfigure}
\usepackage{booktabs}
\usepackage{comment}%

\usepackage{hyperref}

\usepackage[accepted]{icml2024}

\AtBeginDocument{%
    \addtolength{\footskip}{2.0\baselineskip}%
    \fancyfoot[L]{\textit{\textbf{Preprint}}}%
}

\usepackage{amsmath}
\usepackage{amssymb}
\usepackage{mathtools}
\usepackage{makecell}
\usepackage{amsthm,enumitem}

\usepackage{amsmath,amsfonts,bm}

\def\eqref#1{equation~\ref{#1}}

\def\1{\bm{1}}

\def\rva{{\mathbf{a}}}

\def\rvd{{\mathbf{d}}}
\def\rve{{\mathbf{e}}}

\def\rvh{{\mathbf{h}}}

\def\rvm{{\mathbf{m}}}

\def\rvv{{\mathbf{v}}}

\def\rvx{{\mathbf{x}}}
\def\rvy{{\mathbf{y}}}

\def\rmI{{\mathbf{I}}}

\def\rmM{{\mathbf{M}}}

\def\rmP{{\mathbf{P}}}

\def\rmV{{\mathbf{V}}}

\def\rmX{{\mathbf{X}}}

\DeclareMathAlphabet{\mathsfit}{\encodingdefault}{\sfdefault}{m}{sl}
\SetMathAlphabet{\mathsfit}{bold}{\encodingdefault}{\sfdefault}{bx}{n}

\def\gC{{\mathcal{C}}}
\def\gD{{\mathcal{D}}}

\def\gG{{\mathcal{G}}}

\def\gM{{\mathcal{M}}}
\def\gN{{\mathcal{N}}}

\def\gP{{\mathcal{P}}}

\def\sR{{\mathbb{R}}}

\def\sY{{\mathbb{Y}}}

\DeclareMathOperator*{\argmax}{arg\,max}

\setlist{nolistsep,leftmargin=*}
\usepackage[capitalize,noabbrev]{cleveref}

\theoremstyle{plain}

\theoremstyle{definition}

\theoremstyle{remark}

\usepackage[textsize=tiny]{todonotes}

\newcommand{\ourwork}{\textsc{TAGMol}}

\usepackage{multirow}

\icmltitlerunning{\ourwork{}: Target-Aware Gradient-guided Molecule Generation}

\begin{document}

\twocolumn[
\icmltitle{\ourwork{}: Target Aware Gradient-guided Molecule Generation}

\icmlsetsymbol{equal}{*}

\begin{icmlauthorlist}
\icmlauthor{Vineeth Dorna}{equal,mai,cse_umass}
\icmlauthor{D. Subhalingam}{equal,mai}
\icmlauthor{Keshav Kolluru}{mai}
\icmlauthor{Shreshth Tuli}{mai,happening}
\icmlauthor{Mrityunjay Singh}{mai}
\icmlauthor{Saurabh Singal}{mai}
\icmlauthor{N. M. Anoop Krishnan}{civil_iitd}
\icmlauthor{Sayan Ranu}{cse_iitd}
\end{icmlauthorlist}

\icmlaffiliation{mai}{Molecule AI, New Delhi, India}
\icmlaffiliation{cse_iitd}{Department of Computer Science and Engineering, Indian Institute of Technology Delhi, New Delhi, India}
\icmlaffiliation{civil_iitd}{Department of Civil Engineering, Indian Institute of Technology Delhi, New Delhi, India}
\icmlaffiliation{cse_umass}{Manning College of Information and Computer Sciences, University of Massachusetts Amherst, Massachusetts, USA}
\icmlaffiliation{happening}{Happening Technology Ltd, London, England}

\icmlcorrespondingauthor{Vineeth Dorna}{vineeth.dorna@moleculeai.com}
\icmlcorrespondingauthor{D. Subhalingam}{subhalingam.d@moleculeai.com}

\icmlkeywords{Structure-Based Drug Design, Drug-Target Binding, Property prediction, Conditional Diffusion, Property guidance}

\vskip 0.3in
]

\printAffiliationsAndNotice{\icmlEqualContribution} %

\begin{abstract}
\vspace{-0.05in}

3D generative models have shown significant promise in \textit{structure-based drug design (SBDD)}, particularly in discovering ligands tailored to specific target binding sites. Existing algorithms often focus primarily on ligand-target binding, characterized by binding affinity. Moreover, models trained solely on target-ligand distribution may fall short in addressing the broader objectives of drug discovery, such as the development of novel ligands with desired properties like drug-likeness, and synthesizability, underscoring the multifaceted nature of the drug design process.
To overcome these challenges, we decouple the problem into molecular generation and property prediction. The latter synergistically \textit{guides} the diffusion sampling process, facilitating guided diffusion and resulting in the creation of meaningful molecules with the desired properties. We call this guided molecular generation process as \ourwork{}.
Through experiments on benchmark datasets, \ourwork{} demonstrates superior performance compared to state-of-the-art baselines, achieving a $22\%$ improvement in average Vina Score and yielding favorable outcomes in essential auxiliary properties. This establishes \ourwork{} as a comprehensive framework for drug generation.\looseness=-1

\end{abstract}

\vspace{-0.1in}
\section{Introduction}
\label{introduction}

The presence of molecular data featuring 3D spatial information has paved the way for the complex field of \textit{structure-based drug design (SBDD)}~\cite{anderson2003process}. The advent of generative AI for molecules has accelerated the rational drug-design---in contrast to the traditional Edisonian trial-and-error approach---with the goal of creating drug-like molecules in 3D space that effectively bind to specific targets. Specifically, deep generative models, such as those proposed by \citet{luo20213d, liu2022graphbp, peng2022pocket2mol}, autoregressively generate atoms and bonds, while \citet{zhang2023molecule} generate motifs. Despite their progress, the performance of autoregressive models heavily relies on the order of generation, as they condition on previously generated atoms, which can lead to error propagation. Alternatively, diffusion models \cite{sohl2015deep, ho2020denoising} overcome this limitation by conditioning upon all the atoms simultaneously and there by efficiently generating realistic molecules that demonstrate stronger binding affinities with their intended targets.
\looseness=-1

The effectiveness of these generative models is heavily reliant on how well training datasets --- consisting of protein-ligand complexes --- align with desired properties such as binding affinity, drug-like topological features, synthetic accessibility (ease of synthesis), to name a few. For instance, CrossDocked 2020~\cite{francoeur2020three}, a widely-used training dataset for SBDD tasks, predominantly includes complexes with moderate binding affinities. Consequently, models trained solely on such datasets may yield sub-optimal molecules, intrinsically tying their success closely to the dataset's quality. Moreover, drug generation is a multi-faceted process that encompasses not only binding affinity but also a range of other desired properties. Compiling a comprehensive dataset that encompasses a broad range of desired properties poses significant challenges, particularly associated with the high computational costs of assessing a suite of properties. Additionally, refining datasets to meet specific quality constraints can drastically reduce the volume of usable training data. As the number of constraints grows, the likelihood of finding samples that satisfy all these criteria becomes increasingly difficult, further exacerbating the situation. Consequently, generative models trained on lower-quality, but high-volume, data may unintentionally capture suboptimal signals, leading to diminished performance in the context of drug discovery. This situation prompts an exploration into how, during the denoising phase of a generative model, we can effectively introduce desired signals while ensuring meaningful reconstruction.\looseness=-1

To address these challenges, we introduce \ourwork{}-- (\underline{T}arget-\underline{A}ware \underline{G}radient-guided \underline{Mol}ecule Generation), wherein we decouple molecular generation and property prediction. %
We start by training a time-dependent \textit{guide} model that predicts properties from inputs with noise levels similar to those in the base diffusion model. Crucially, we turn the challenge of using inferior quality data, i.e., the property of interest is well spread with the inclusion of sub-optimal values in the dataset, to our strength for robust \textit{guide} training. Inspired by classifier guidance in diffusion models \cite{dhariwal2021diffusion}, we use the gradient of \textit{guide} to direct the latent space during the diffusion sampling process, ensuring the reconstructed molecules possess the targeted properties. In the sampling phase, we harness the strengths of both the generative model and the \textit{guide}. This interactive dynamic enables us to explore regions with superior properties while simultaneously denoising the latent space to generate diverse molecules. 
While gradient guidance is a well-explored concept in the drug discovery domain, many existing methods overlook target awareness \cite{bao2022equivariant} or fail to integrate 3D structure \cite{eckmann2022limo, stanton2022accelerating, lee2023MOOD}. In contrast, our approach seeks to simultaneously optimize for both target-aware and molecular properties in 3D space.
Our method marks a significant advancement in end-to-end molecular generation, reducing the reliance on post-optimization techniques. Furthermore, to address real-world scenarios where multiple property constraints exist, we train separate \textit{guides} for each property, subsequently employing them to steer the diffusion process effectively.\looseness=-1

Overall, the key contributions of our work are as follows.
\vspace{-0.1in}
\begin{itemize}[itemsep=0pt]
\item \textbf{Reformulation of the drug-discovery problem:} We reformulate the problem of drug generative modeling moving beyond the myopic lens of optimizing binding activity. The need to optimize other properties of interest, even when these signals are not adequately present in the train set, is explicitly coded into our problem formulation.
\item \textbf{Algorithm design:} We design a novel generative process, called \ourwork{}, which jointly leverages the signals from two different components: an \textit{SE(3) equivariant graph diffusion model} to mimic geometries of the train set, and a \textit{multi-objective guide model}, empowered by an SE(3) invariant GNN, to steer the exploration region of diffusion sampling towards the property of interest by leveraging gradients.\looseness=-1
\item \textbf{Rigorous empirical evaluation:} 
We demonstrate that our model achieves $22\%$ improvement in average Vina Score, all the while being guided by considerations of binding affinity and crucial pharmacological properties such as QED and SA.
\end{itemize}

\begin{figure*}[!htp]
  \centering
  \includegraphics[width=5.9in]{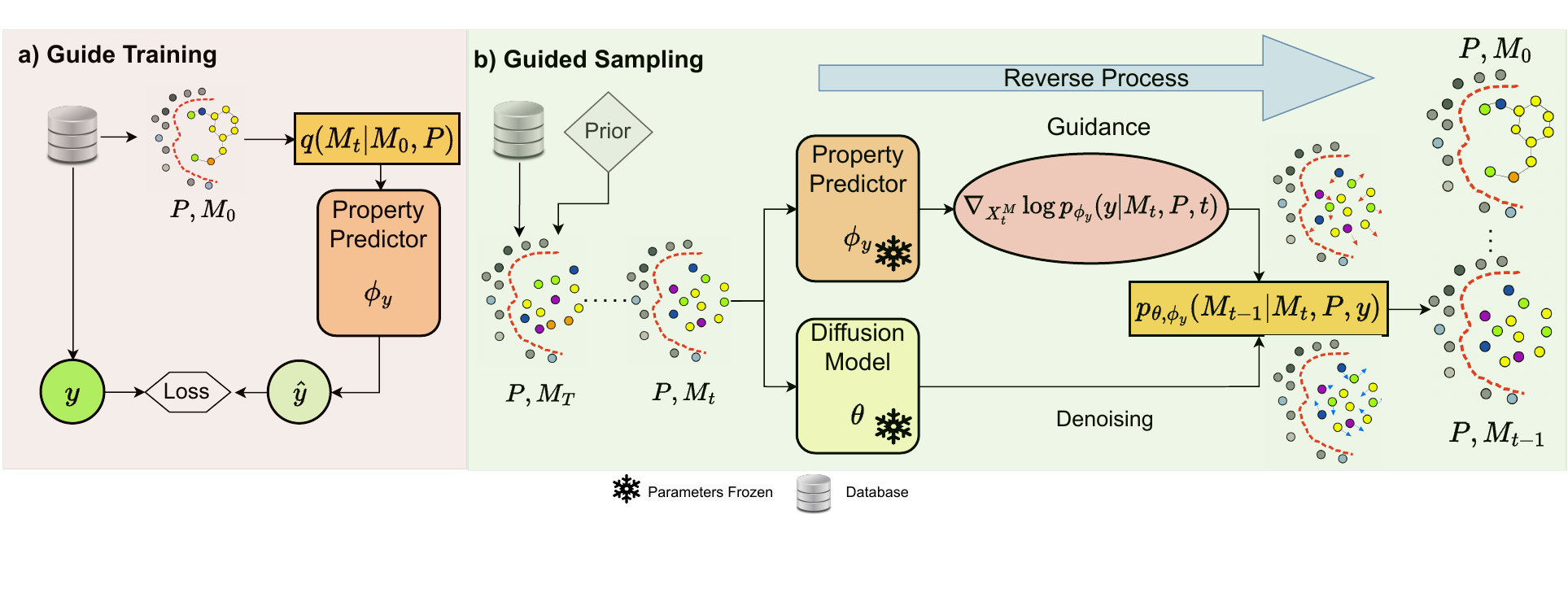}
  \vspace{-0.4in}
  \caption{Overview of \ourwork{}. (a) Training a property-oriented \textit{guide} using existing data. (b) Utilizing the trained \textit{guide} and \textit{diffusion} model to steer the diffusion sampling process towards the optimal regions of the property of interest.}
  \label{fig:framework}
  \vspace{-0.2in}
\end{figure*}

\section{Preliminary}
In this section, we introduce the preliminary concepts central to our work and formulate the problem of gradient-guided molecular generation. All notations introduced in this section are summarized in Table~\ref{tab:notations} in the Appendix.
\vspace{-0.1in}
\subsection{Problem Statement}
\label{sec:prob}
\vspace{-0.05in}
From the standpoint of generative models, SBDD task involves the creation of ligand molecules with the ability to bind effectively to a specified protein binding site. A protein binding site is characterized by a set of atoms, denoted by $\gP = \{ \left(\rvx^{\gP}_{i}, \rvv^{\gP}_{i}\right) \}_{i=1}^{N_{\gP}}$, where $N_{\gP}$ represents the number of protein atoms, $\rvx^{\gP}_{i} \in \sR^{3}$ represents the 3D coordinates of the protein atom, and $\rvv^{\gP}_{i} \in \sR^{N_f}$ represents protein atom features such as element types and amino acid types, with ${N_f}$ representing the number of such features. We aim to jointly optimize binding affinity and desired pharmacological properties (denoted as $\sY$), by generating prospective ligand molecules $\gM = \{ \left(\rvx^{\gM}_{i}, \rvv^{\gM}_{i}\right) \}_{i=1}^{N_{\gM}}$, for a given protein $\gP$. Here $\rvx^{\gM}_{i} \in \sR^{3}$ and $\rvv^{\gM}_{i} \in \sR^{K}$ represents the atom coordinates and atom types of a ligand molecule, respectively, with $K$ representing the number of such features to represent atom types. The variable $N_{{\gM}}$ signifies the number of atoms in the ligand molecule, which can be sampled during inference utilizing either an empirical distribution~\cite{hoogeboom2022equivariant,guan2023d, guan2023decompdiff} or predicted through a neural network~\cite{liu2022graphbp}. 

To simplify, in matrix representation the ligand molecule is denoted as $\rmM = [\rmX^{\gM}, \rmV^{\gM}]$, where  $\rmX^{\gM} \in \mathbb{R}^{N_{\gM} \times 3}$ and $\rmV^{\gM} \in \mathbb{R}^{N_{\gM} \times K}$  and the protein is denoted as $\rmP = [\rmX^{\gP}, \rmV^{\gP}]$, where  $\rmX^{\gP} \in \mathbb{R}^{N_{\gP} \times 3}$ and $\rmV^{\gP} \in \mathbb{R}^{N_{\gP} \times N_{f}}$. 

\vspace{-0.1in}
\subsection{Diffusion models for Target-Aware Generation}
\label{sec:diffusion}
\vspace{-0.05in}
As delineated in previous works on diffusion-based target-aware molecular generations \cite{guan2023d, guan2023decompdiff}, the process involves two phases: \textit{noise injection}, also termed as forward diffusion, and \textit{denoising} (backward diffusion).

\textbf{Noise injection:} This phase involves a gradual injection of Gaussian noise for co-ordinates and uniform noise for categorical data through a Markov chain. This noise addition is uniquely applied to the ligand molecule, excluding the protein in the diffusion process. In this context, the atom positions and types of the ligand molecule at time step $t$ are represented as $\rmX_{t}^{\gM}$ and $\rmV_{t}^{\gM}$ respectively. The diffusion forward transition is defined by the following equations:
\begin{equation}
\begin{small}
\begin{aligned}
\label{eq:detail_diff_one_step}
q\left(\rmM_t | \rmM_{t-1}, \rmP\right) &= \gN\left(\rmX^{\gM}_{t}; \sqrt{1-\beta_t}\rmX^{\gM}_{t-1}, \beta_t \rmI\right) \\
&\quad \cdot \gC\left(\rmV^{\gM}_{t} | \left(1-\beta_t\right) \rmV^{\gM}_{t-1}  + \beta_t / K\right)
\end{aligned}
\end{small}
\end{equation}
\begin{equation}
\begin{small}
\begin{aligned}
\label{eq:detail_diff_big_setp}
q\left(\rmM_t | \rmM_{0}, \rmP\right) &= \gN\left(\rmX^{\gM}_{t}; \sqrt{\bar{\alpha_t}} \rmX^{\gM}_{0}, \left(1-\bar{\alpha_t}\right) \rmI\right) \\
&\quad \cdot \gC\left(\rmV^{\gM}_{t} | \bar{\alpha_t} \rmV^{\gM}_{0}  + \left(1-\bar{\alpha_t}\right) / K\right)
\end{aligned}
\end{small}
\end{equation}
where $\gN$ and $\gC$  denotes the normal and categorical distribution respectively while $\beta_1, ..., \beta_T$ represents the variance schedules. The corresponding posteriors are analytically derived as follows:
\vspace{-0.1in}
\begin{equation}
\begin{small}
\begin{aligned}
\label{eq:posterior}
q\left(\rmM_{t-1} | \rmM_{0}, \rmM_{t}, \rmP\right) &= \gN\left(\rmX^{\gM}_{t-1}; \Tilde{\mu}_t\left(\rmX^{\gM}_{t}, \rmX^{\gM}_{0}\right), \Tilde{\beta_t} \rmI\right) \\
&\cdot \gC\left(\rmV^{\gM}_{t-1} | \Tilde{\bm{c}_t}\left(\rmV^{\gM}_{t}, \rmV^{\gM}_{0}\right)\right)
\end{aligned}
\end{small}
\end{equation}
where, \\
 $\Tilde{\mu}_t\left(\rmX^{\gM}_{t}, \rmX^{\gM}_{0}\right) = \frac{\sqrt{\bar\alpha_{t-1}} \beta_t}{1 - \bar\alpha_t} \rmX^{\gM}_{0} + \frac{\sqrt{\alpha_t}\left(1 - \bar\alpha_{t-1}\right)}{1 - \bar\alpha_t}\rmX^{\gM}_{t}$, \\
 $\Tilde{\beta}_t = \frac{1 - \bar\alpha_{t-1}}{1 - \bar\alpha_t} \beta_t$, $\alpha_t = 1-\beta_t$, 
 $\bar{\alpha}_t = \Pi_{s=1}^t \alpha_s$, 
 $\Tilde{c}_t\left(\rmV^{\gM}_{t}, \rmV^{\gM}_{0}\right) = \bm{c}^\star / \sum_{k=1}^K c_k^\star$, and $\bm{c}^\star\left(\rmV^{\gM}_{t}, \rmV^{\gM}_{0}\right) = [\alpha_t\rmV^{\gM}_{t} + \left(1 - \alpha_t\right) / K] \odot [\bar\alpha_{t-1}\rmV^{\gM}_{0} + \left(1 - \bar\alpha_{t-1}\right) / K]$.

In practical applications, it is recognized that the schedules \(\beta_t\) for coordinates and categories may differ. However, for the sake of simplicity in this context, they are uniformly represented.\\
\textbf{Denoising phase:} In the generative process, a neural network parameterized by $\theta$ learns to recover $\rmM_0$ by iteratively denoising $\rmM_T$. During reverse process, $\rmM_{0}$ is approximated using $\theta$ and $\rmM_{t-1}$  by predicting $\widehat{\rmM}_{0|t} = [\widehat{\rmX}^{\gM}_{0|t}, \widehat{\rmV}^{\gM}_{0|t}]$  at time step $t$ and $\rmM_{t-1}$ is sampled as follows:
 \vspace{-0.10in}
\begin{equation}
\begin{small}
\begin{aligned}
\label{eq:detail_revdiff}
p_{\theta}\left(\rmM_{t-1} | \rmM_t, \rmP\right) &= q\left(\rmM_{t-1} | \widehat{\rmM}_{0|t}, \rmM_{t}, \rmP\right) \\
&= \gN\left(\rmX^{\gM}_{t-1}; \Tilde{\mu}_{\theta}\left(\rmM_t, \rmP, t\right) , \Tilde{\beta_t} \rmI\right) \\
&\quad \cdot \gC\left(\rmV^{\gM}_{t-1} | \Tilde{\bm{c}_{\theta}}\left(\rmM_t, \rmP, t\right)\right) \\
&= \gN\left(\rmX^{\gM}_{t-1}; \Tilde{\mu}_t\left(\rmX^{\gM}_{t}, \hat{\rmX}^{\gM}_{0|t}\right) , \Tilde{\beta_t} \rmI\right) \\
&\quad \cdot \gC\left(\rmV^{\gM}_{t-1} | \Tilde{\bm{c}_t}\left(\rmV^{\gM}_{t}, \hat{\rmV}^{\gM}_{0|t}\right)\right)
\end{aligned}
\end{small}
\end{equation}
\textbf{Training:} In line with the principles outlined in the variational auto-encoder \cite{kingma2013auto}, the model undergoes training through the optimization of variational bound on the negative log-likelihood. Given that both $q\left(\rmX_{t-1} | \rmM_{0}, \rmM_{t}, \rmP\right)$ and $p_{\theta}\left(\rmX_{t-1} | \rmM_t, \rmP\right)$ are Gaussian distributions, the Kullback-Leibler (KL) divergence for the atom coordinates is expressed in a closed-form equation:
\vspace{-0.1in}
\begin{equation}
\begin{small}
\begin{aligned}
L_{t-1}^{\left(x\right)} &= \frac{1}{2\Tilde{\beta}_t} \| \Tilde{\mu}_t\left(\rmX^{\gM}_{t}, \rmX^{\gM}_{0}\right) - \Tilde{\mu}_{\theta}\left([\rmX^{\gM}_{t}, \rmV^{\gM}_{t}], \rmP, t\right)\|^2 + C \\
&= \gamma_t \|\rmX^{\gM}_{0} - \widehat{\rmX}^{\gM}_{0|t} \|^2 + C
\end{aligned}
\end{small}
\end{equation}
where $\gamma_t = \frac{\bar\alpha_{t-1} \beta_t^2}{2\sigma_t^2\left(1 - \bar\alpha_t\right)^2}$ and $C$ is a constant. As recommended by \cite{ho2020denoising} and \cite{guan2023d}, training the model using an unweighted Mean Squared Error (MSE) loss, by setting $\gamma_t=1$, leads to enhanced performance. Regarding the atom-type loss, the KL divergence of categorical distributions is computed in the following manner:
\begin{equation}
\begin{aligned}
\label{eq:unweight_pos_loss}
L_{t-1}^{\left(v\right)} = \sum_{k=1}^{K} \bm{c}\left(\rmV^{\gM}_{t}, \rmV^{\gM}_{0}\right)_k \log \frac{\bm{c}\left(\rmV^{\gM}_{t}, \rmV^{\gM}_{0}\right)_k}{\bm{c}\left(\rmV^{\gM}_{t}, \widehat{\rmV}^{\gM}_{0|t}\right)_k}
\end{aligned}
\end{equation}

The final loss is a weighted sum of atom coordinate loss and atom type loss, which is expressed as $L = L_{t-1}^{\left(x\right)} + \lambda L_{t-1}^{\left(v\right)}$.

\section{\ourwork}
\label{sec:methods}
In this work, we improve upon \textit{Conditional Diffusion models} with a particular emphasis on utilizing protein pockets as a conditioning factor for generating ligand molecules \cite{guan2023d, guan2023decompdiff}. The effectiveness of such models in practical scenarios can be hindered if the conditioning signal is overlooked or weakened, an issue that becomes more pronounced with datasets of inferior quality. %
Models that are exclusively trained to maximize the likelihood of protein-ligand complexes in such datasets naturally inherit the same quality issues. %
To counteract this, \ourwork{} employs a strategic approach to explicitly integrate additional conditioning signals learned over a set of binding and desirable pharmacological properties $\mathbb{Y}$ (Recall \S~\ref{sec:prob}) during the model's denoising phase. This tactic aligns with the method proposed by \cite{dhariwal2021diffusion}, which involves conditioning a pre-trained diffusion model using classifier gradients. In particular, for a property $\rvy \in \mathbb{Y} $, we train a regressor (or classifier as appropriate) $p_{\phi}(\rvy | \mathbf{M}_t, \mathbf{P}, t)$ on noisy molecule $\mathbf{M}_t$ and then use the gradients $\nabla_{\rmX^{\gM}_t} p_{\phi}(\rvy | \mathbf{M}_t, \mathbf{P}, t)$ to guide the diffusion sampling process towards the desirable properties encoded in $\rvy$.\looseness=-1

Fig~\ref{fig:framework} presents the pipeline of \ourwork{}. \ourwork{} consists of two key components. The first component is \textit{Guide Training}, introduced in \S~\ref{subsec:prop_guide}, where we present our \textit{SE(3) Invariant} GNN architecture and its modeling objectives, establishing the foundation for our approach. The goal of this step is to assess how well a molecule conforms to the desirable properties $\mathbb{Y}$. The second component, discussed in \S~\ref{subsec:single_objective}, is \textit{Guided Sampling}. Here, we elaborate on how a trained guide model efficiently steers the diffusion sampling process towards specific regions of interest, emphasizing desired properties. Our study goes further to demonstrate the concurrent optimization of multiple properties, as detailed in \S~\ref{subsec:multi_objective}. This multifaceted approach more holistically addresses the various aspects of molecular design, ultimately simplifying the efforts of chemists devote to the lead optimization process.
\looseness=-1

\vspace{-0.1in}
\subsection{Property Guide}
\label{subsec:prop_guide}
\vspace{-0.05in}
We approach the input space as a 3D point cloud system where we build a $k$-nearest neighbors ($k$-NN) graph $\gG$ by representing ligand and protein atoms as nodes, and each atom is connected to its $k$-nearest neighbors. For properties such as binding affinity, including protein atoms within the graph is critical; however, the inclusion of protein atoms can be omitted for properties solely dependent on the ligand. We parameterize our guide model using an Invariant GNN $\phi_y$, which is trained to predict property $\rvy \in \sY$ given a noisy input $\rmM_t$, protein $\rmP$ and time $t$. Later in the denoising phase, the gradients of $\phi_y$ are used to direct exploration in regions of interest. %
Given the graph representation $\gG$ at diffusion time step $t$, we define the GNN convolution layer as follows:
 \vspace{-0.2in}
\begin{equation}
\begin{small}
\begin{aligned}
\label{eq:guide_net_conv}
    \rvd_{i,j} &= \gD( \|\rvx_{i, t} - \rvx_{j, t} \|^2) \\
    \rvm_{i,j} &= \phi^{m}_{\rvy}(\rvh^{l}_{i,t}, \rvh^{l}_{j,t}, \rvd_{i,j}, \rva_{i,j}) \\
    \rve_{i,j} &= \phi^{e}_{\rvy} (\rvm_{i,j}) \\
    \rvm_{i} &= \sum_{j \in \gN(i)} \rve_{i,j} \rvm_{i,j} \\
    \rvh^{l+1}_{i,t} &= \phi^{h}_{\rvy} (\rvh^{l}_{i,t}, \rvm_{i}) \\
    \rvh^{0}_{i,t} &= \begin{cases}
        \phi_{\rvy}^{p}(\rvv^{\gP}_{i}) & \text{if } i \in \gP\\
        \phi_{\rvy}^{l}([\rvv^{\gM}_{i,t}, \tau]) & \text{if } i \in \gM
    \end{cases}
\end{aligned}
\end{small}
\end{equation}
where $\rvh^{l}_{i,t}  \in \sR^d$ represents the SE(3)-invariant hidden representation of protein and ligand atoms after $l$ layers;  $\rvx_{i, t}, \rvx_{j, t} \in \sR^3$ represents hidden representation of protein and ligand atoms. $\rvd_{i,j}$ represents the distance embedding and $\rva_{i,j}$ represents the edge attributes in the graph. Invariance of $\rvh^{l}_{i,t}$ stems from the fact that we take the L2-distance between the atom representations. $\gN(i)$ stands for the set of neighbors for atom $i$. 
$\tau$ represents the time embedding to make the model aware of noise at time step $t$.
$\phi^{m}_{\rvy}$,  $\phi^{e}_{\rvy}$, $\phi^{h}_{\rvy}$, $\phi^{p}_{\rvy}$ and $\phi^{l}_{\rvy}$ are  Multi-Layer Perceptrons (MLP) where as $\gD$ is a distance encoder.
\looseness=-1

Once we get the final hidden states $\rvh^{L}_{i,t}$, we predict the final property using an MLP layer as:
 \vspace{-0.1in}
\begin{equation}
\begin{aligned}
\label{eq:guide_net_prop}
\hat{\rvy} &= \phi^{f}_{\rvy} \left(\sum_{i \in \gM} \rvh^{L}_{i,t}\right)
\end{aligned}
\end{equation}
While equivariance is essential for generative models to maintain the consistency of the probability $p(\rmM_0 | \rmP)$ against \textit{SE(3)} transformations in protein-ligand complexes, it is also vital that a scalar property predicted from \textit{guide} remains independent of \textit{SE(3)} transformations. Hence, the GNN in guide is SE(3) invariant. This modeling approach introduces a valuable inductive bias, which is particularly beneficial in the generation of 3D objects, such as molecules \cite{kohler2020equivariant, satorras2021enf, xu2022geodiff, hoogeboom2022equivariant, guan2023d, guan2023decompdiff}.

In our approach,  diverging from the classifier-guided methodology \cite{dhariwal2021diffusion} in diffusion processes where the classifier directly estimates probabilities, we train our model to regress specific properties. The probability modeling is articulated through a normal distribution as:
\begin{equation}
\begin{aligned}
\label{eq:guide_net_prob}
    p_{\phi_y}(\rvy | \rmM_t, \rmP) = \gN(\rvy, \phi_\rvy(\rmM_t, \rmP, t), \rmI)
\end{aligned}
\end{equation}
The training of these models is oriented toward minimizing the Negative log-likelihood ($NLL$), which effectively simplifies the final objective to minimizing the Root Mean Square Error (RMSE) between $\rvy$ and $\phi_\rvy(\rmM_t, \rmP, t)$.
\begin{equation}
\begin{aligned}
\label{eq:guide_net_objective}
    NLL &= - \mathbb{E}_{p(\rmP, \rmM_{0:T})} \sum_{t=0}^{T} \log(p_{\phi_y}(\rvy | \rmM_t, \rmP, t)) \\
    &= \mathbb{E}_{p(\rmP, \rmM_{0:T})} \sum_{t=0}^{T} \frac{(\rvy - \phi_\rvy(\rmM_t, \rmP, t))^2}{2} \\
\end{aligned}
\end{equation}
Although we designed our guide for regression tasks, it is versatile enough to be adapted for classification tasks or other suitable applications.
\vspace{-0.1in}
\subsection{Single Objective Guidance}
\label{subsec:single_objective}
\vspace{-0.05in}
In this section, we first design the framework to obtain guidance from a single property. Subsequently, in \S~\ref{subsec:multi_objective}, we extend to multi-objective guidance.

Departing from the methodologies outlined in prior studies \cite{guan2023d, guan2023decompdiff}, our generative model distinctively conditions on both the protein pocket and property to be guided. Remarkably, without retraining the diffusion model, we guide the diffusion sampling process by shifting coordinates. This approach empowers the model to consistently condition on optimal property values during the denoising phase. By steering the denoising process towards an optimal property, it significantly enhances the model's capability to generate molecules with desired characteristics.\looseness=-1

As demonstrated in \cite{dhariwal2021diffusion}, efficiently sampling $\rmM_{t-1}$ for each denoising transition can be adequately achieved by: 
\vspace{-0.05in}
\begin{equation}
\begin{aligned}
\label{eq:single_guide_prob}
p_{\theta,\phi_{\rvy}}(\rmM_{t-1} | \rmM_t, \rmP, \rvy) = Z & p_{\theta} (\rmM_{t-1} | \rmM_t, \rmP) \\
& \cdot p_{\phi_{\rvy}} (\rvy | \rmM_{t-1}, \rmP, t-1)
\end{aligned}
\end{equation}
where $Z$ is a normalizing constant. However, direct sampling from this distribution is intractable. We approximate the sampling process via perturbation to a Gaussian distribution as per prior work~\cite{sohl2015deep,dhariwal2021diffusion}. Thus, we use perturbation to $p_{\theta} (\rmX_{t-1} | \rmM_t, \rmP)$ and sample $\rmX^{\gM}_{t-1}$ as:
\begin{equation}
\begin{small}
\begin{aligned}
\label{eq:single_guide_samp}
\rmX^{\gM}_{t-1} &\sim \gN(\Tilde{\mu}_{\theta}(\rmM_t, \rmP, t)+ s \Tilde{\beta} \nabla_{{\rmX}^{\gM}_t} \log p_{\phi_{\rvy}} (\rvy | \rmM_{t}, \rmP, t), \Tilde{\beta}\rmI) \\
\rmV^{\gM}_{t-1} &\sim \gC\left(\Tilde{\bm{c}_{\theta}}\left(\rmM_t, \rmP, t\right)\right)
\end{aligned}
\end{small}
\end{equation}
In this context, the parameter $s$, denoting the guidance strength, plays a crucial role in prioritizing $\rmM_t$ with optimal property by sampling it from the updated distribution $\propto p_{\theta} (\rmM_{t-1} | \rmM_t, \rmP)  \left( p_{\phi_{\rvy}} (\rvy | \rmM_{t-1}, \rmP, t) \right)^s$. Fine-tuning of $s$ is essential during the optimization process to improve property prediction while maintaining the effectiveness of denoising within the diffusion model.

The incorporation of discrete variables such as atom types $\rmV^{\gM}_{t}$ in the diffusion process poses challenges in directly applying explicit guidance through gradients. However, in our approach, while we explicitly provide guidance for coordinates $\rmX^{\gM}_{t}$, the guidance for discrete variables operates implicitly. More specifically, at diffusion time step $t$, the denoised ligand atom types $\rmV^{\gM}_{t-1}$ are influenced by both $\rmX^{\gM}_{t}$ and $\rmV^{\gM}_{t}$. Therefore, by providing guidance for $\rmX^{\gM}_{t}$, the generative model is encouraged to denoise $\rmV^{\gM}_{t-1}$ for the optimized $\rmX^{\gM}_{t}$, effectively making the guidance implicit for the atom types.

\vspace{-0.05in}
\subsection{Multi Objective Guidance}
\label{subsec:multi_objective}
\vspace{-0.05in}
We now generalize our objective to holistically enhance various desired properties, collectively denoted as $\sY$, in the molecules we generate. At each denoising step $t$, we condition on $\sY$ and sample $\rmM_{t-1}$ according to the probability distribution similar to Equation \ref{eq:single_guide_prob} as :
 \vspace{-0.05in}
\begin{equation}
\begin{aligned}
\label{eq:multi_guide_prob_naive}
p_{\theta,\phi_{\sY}}(\rmM_{t-1} | \rmM_t, \rmP, \sY) = Z & p_{\theta} (\rmM_{t-1} | \rmM_t, \rmP) \\
& \cdot p_{\phi_{\sY}} (\sY | \rmM_{t-1}, \rmP, t)
\end{aligned}
\end{equation}
Assuming all the properties $\rvy \in \sY$ are conditionally independent given $\rmM_{t-1}$ and $\rmP$, Eq.~\ref{eq:multi_guide_prob_naive} is factorized as follows:
\vspace{-0.1in}
\begin{equation}
\begin{small}
\begin{aligned}
\label{eq:multi_guide_prob}
p_{\theta,\phi_{\sY}}(\rmM_{t-1} | \rmM_t, \rmP, \sY) = Z & p_{\theta} (\rmM_{t-1} | \rmM_t, \rmP) \\
& \cdot \prod_{\rvy \in \sY} p_{\phi_{\rvy}} (\rvy | \rmM_{t-1}, \rmP, t)
\end{aligned}
\end{small}
\end{equation}
Thus we train a set of models $\phi_{\sY}= \{ \phi_{\rvy} :  \forall \rvy \in \sY \}$ independently and sample $\rmM_{t-1}$ using a similar approximated posterior distribution in Equation \ref{eq:single_guide_samp} as:
\begin{equation}
\begin{aligned}
\label{eq:multi_guide_samp}
\rmX^{\gM}_{t-1} &\sim \gN(\Tilde{\mu}_{\theta} (\rmM_t, \rmP, t) + \delta, \Tilde{\beta}\rmI) \\
\rmV^{\gM}_{t-1} &\sim \gC\left(\Tilde{\bm{c}_{\theta}}\left(\rmM_t, \rmP, t\right)\right)
\end{aligned}
\end{equation}
where 
 \vspace{-0.3in}
\begin{equation}
\begin{aligned}
\label{eq:multi_guide_term}
\delta = \sum_{\rvy \in \sY} s_{\rvy} \Tilde{\beta} \nabla_{{\rmX}^{\gM}_t} \log p_{\phi_{\rvy}} (\rvy | \rmM_{t}, \rmP, t) 
\end{aligned}
\end{equation}
and $s_{\rvy}$ represents the guidance strength for property $\rvy \in \sY$.
\vspace{-0.1in}
\section{Experiments}
\vspace{-0.05in}
In this section, we benchmark \ourwork{} against state-of-the-art molecular generative models and establish that gradient guidance leads to superior performance across an array of important metrics assessing binding and pharmacological properties.\looseness=-1

\vspace{-0.1in}
\subsection{Datasets} 
\label{sec:datasets}
\vspace{-0.05in}
\ourwork{} is trained and evaluated on the CrossDocked2020 dataset~\cite{francoeur2020three}, consistent with the approaches outlined in \citet{luo20213d, peng2022pocket2mol, guan2023d, guan2023decompdiff}. The dataset originally comprised of 22.5 million docked binding complexes, which undergo a rigorous refinement process. This refinement narrows down the dataset to better-quality docking poses, ensuring that the root-mean-square deviation (RMSD) between the docked pose and the ground truth is less than 1Å, and that the protein sequences exhibit less than 30\% identity. From this refined pool, 100,000 protein-ligand pairs are selected for the training set. For the evaluation, we employ a separate set of 100 proteins that are distinct from those in the training dataset. To ensure a fair comparison with our baseline methods, we adhere to identical data splits for training our \textit{guide} models and evaluating our overall method.
\vspace{-0.1in}
\subsection{Baselines}
\vspace{-0.05in}
We benchmark against state-of-the-art baselines in the realm of structure-based drug design (SBDD). This includes liGAN \cite{ragoza2022generating}, which leverages a conditional variational autoencoder (CVAE) and is trained on a grid representation of atomic densities in protein-ligand structures. Additionally, we consider AR \cite{luo20213d}, 
and Pocket2Mol \cite{peng2022pocket2mol}, GNN-based methods both of which employ autoregressive frameworks to generate 3D molecular atoms by conditioning on the protein pocket and previously generated atoms. Furthermore, our comparison extends to recent diffusion-based approaches such as TargetDiff \cite{guan2023d} and DecompDiff \cite{guan2023decompdiff}, which have set new benchmarks in the non-autoregressive generation of atom coordinates and types. DecompDiff enhances TargetDiff by integrating bond considerations and introducing decomposed priors for the ligand's arms and scaffolds.  For a comprehensive overview of prior works and additional references, please refer to the \S~\ref{sec:related_work} of Appendix.

\subsection{Metrics}
\label{sec:metrics}
To evaluate the quality of molecules generated by \ourwork{} and the baselines, we adopt a multi-faceted assessment strategy encompassing molecular properties, their conformation, and binding affinity with the target. 

\subsubsection{Molecular Properties}
\textbf{QED (Quantitative Estimate of Druglikeness).} This metric evaluates a molecule's drug-likeness \cite{bickerton2012quantifying} by reflecting the typical distribution of molecular properties in successful drug candidates.  \\
\textbf{SA (Synthetic Accessibility).} SA assesses the ease with which a molecule can be synthesized \cite{ertl2009estimation}, serving as a vital indicator of its practical manufacturability in a laboratory or industrial setting. \\
\textbf{Diversity.} Diversity is measured as the average pairwise Tanimoto distances \cite{bajusz2015tanimoto, tanimoto1958elementary} among all ligands generated for a specific protein pocket. 
\textbf{Bond Distance Distribution.} We calculate the Jensen-Shannon divergences (JSD) to assess the differences in bond distance distributions between the reference molecules and the generated molecules \cite{guan2023d, guan2023decompdiff}.

\subsubsection{Binding Affinity}

AutoDock Vina\cite{eberhardt2021autodock} is employed to calculate the following metrics: \\
\textbf{Vina Score.} Estimates binding affinity from the 3D pose of generated molecules, where a favorable score suggests strong binding potential. \\
\textbf{Vina Min.} By conducting a local structure minimization prior to affinity estimation, this metric presents a slightly refined perspective on Vina Score. \\
\textbf{Vina Dock.} Incorporating a re-docking process, Vina Dock showcases the optimal binding affinity that can be achieved.\\
\textbf{High Affinity.} This is a comparative metric evaluating the percentage of generated molecules that manifest better binding than a reference molecule for a given protein pocket. 
\subsubsection{Hit Rate}
We report the percentage of molecules meeting \textit{hit criteria} characterized as: QED $\geq$ 0.4, SA $\geq$ 0.5, and Vina Dock $\leq$ -8.18 kcal/mol. These QED, SA thresholds are set in line with the ranges observed in currently marketed drugs \cite{eckmann2022limo}. The threshold for the Vina Dock score is chosen to reflect a binding affinity less than 1$\mu$m. This threshold is commonly used in medicinal chemistry to ensure moderate biological activity \cite{yang2021knowledge, long2022zero,guan2023decompdiff}. Optimizing the hyperparameters in \ourwork{} for \textit{hit rate} allows us to generate higher number of acceptable molecules as per our criteria.

\vspace{-0.1in}
\subsection{Experimental setup}
\label{sec:expt-setup}

\paragraph{Generative Backbone.} For our generative backbone model, we opted for TargetDiff \cite{guan2023d} over DecompDiff \cite{guan2023decompdiff}. The decision was based on TargetDiff's self-contained nature, in contrast to DecompDiff, which relies heavily on external computational tools. Specifically, DecompDiff utilizes AlphaSpace2 \cite{katigbak2020alphaspace} for extracting subpockets, which are potential protein binding sites. In fact, our study demonstrates how our guidance mechanism can effectively replace the computational tools employed in DecompDiff. %

\textbf{Guide Training:}
In our experimental setup, as detailed in \S~\ref{subsec:prop_guide}, we employed a $9$-layer Invariant Graph Neural Network (GNN) as the foundational model for constructing \textit{guides} across various properties. Notably, for QED and SA — properties that are not dependent on the target — we adapted the $k$-Nearest Neighbors ($k$-NN) graph construction methodology, which omits protein atoms into consideration. The $k$ value was specifically tailored to each property: set at 6 for both QED and SA and increased to 32 for the binding affinity property predictor to optimize the model's performance in different contexts. For training our \textit{binding affinity guide}, we used Autodock Vina scores from the CrossDocked2020 dataset \cite{eberhardt2021autodock, francoeur2020three}, and for QED and SA, we calculated scores using the RDKit package \cite{landrum2013rdkit}.
To effectively guide the denoising phase, we train our guide using the same noise as our backbone model, TargetDiff, which applies Gaussian noise to coordinates and uniform noise to categories. Additional training and evaluation details of \textit{guide} can be found in Appendix \S~\ref{sec:train_eval_guide}.

\textbf{Guide Strengths:}
For the single objective guidance process, the optimal guide strength $s_{\rvy}^{opt}$ for each property $\rvy$ is identified through a grid search on small set of 4 targets (see Appendix  \S~\ref{sec:guide-strength-search} for more details).
The configuration that delivers the best value for the intended property in the generated molecules is selected. 
For multi-objective guidance, the optimal guide strength values $s_{\rvy}^{opt}$ are recalibrated using a set of weights $w_{\rvy}$, where $\sum_{\rvy \in \sY} w_{\rvy} = 1$. These modified guide strengths, \(w_{\rvy}s_{\rvy}^{\text{opt}}\), are subsequently utilized to steer property optimization as described in Equation~\ref{eq:multi_guide_samp}.
In our approach, where all properties are considered equally important, we decided to assign equal weights and observed the best \textit{hit rate}. Further analyses on how performance varies with respect to weights $w_{\rvy}$ can be seen in Table \ref{tab:guidance_ablation_all} in Appendix.

\textbf{Evaluation Setup:} Adhering to the setup outlined in our baseline studies, we generated 100 molecules for each of the 100 test proteins and report the metrics  detailed in \S~\ref{sec:metrics}. %

\begin{table*}[!htb]
\centering
\caption{\footnotesize Comparison of various properties between reference molecules and those generated by our model and other baselines. ($\uparrow$) / ($\downarrow$) indicates whether a larger or smaller number is preferable. The first and second-place results are emphasized with bold and underlined text, respectively. Refer to Appendix~\ref{app:posecheck} for evaluation on additional metrics suggested in \cite{harris2023posecheck}.
}
\label{tab:ablation_all}
\vskip 0.1in
\resizebox{\linewidth}{!}{
\begin{tabular}{@{}l|cc|cc|cc|cc|cc|cc|cc|c@{}}
\toprule
Methods &
  \multicolumn{2}{c|}{Vina Score ($\downarrow$)} &
  \multicolumn{2}{c|}{Vina Min ($\downarrow$)} &
  \multicolumn{2}{c|}{Vina Dock ($\downarrow$)} &
  \multicolumn{2}{c|}{High Affinity ($\uparrow$)} &
  \multicolumn{2}{c|}{QED ($\uparrow$)} &
  \multicolumn{2}{c|}{SA ($\uparrow$)} &
  \multicolumn{2}{c|}{Diversity ($\uparrow$)} &
  Hit($\uparrow$)\\ 
 &
  Avg. &
  Med. &
  Avg. &
  Med. &
  Avg. &
  Med. &
  Avg. &
  Med. &
  Avg. &
  Med. &
  Avg. &
  Med. &
  Avg. &
  Med. &
  Rate \%  \\ \midrule
Reference &
  -6.36 &
  -6.46 &
  -6.71 &
  -6.49 &
  -7.45 &
  -7.26 &
  - &
  - &
  0.48 &
  0.47 &
  0.73 &
  0.74 &
  - &
  - &
  21 \\ \midrule
liGAN &
  - &
  - &
  - &
  - &
  -6.33 &
  -6.20 &
  21.1\% &
  11.1\% &
  0.39 &
  0.39 &
  0.59 &
  0.57 &
  0.66 &
  0.67 &
  13.2 \\
AR &
  \underline{-5.75} &
  -5.64 &
  -6.18 &
  -5.88 &
  -6.75 &
  -6.62 &
  37.9\% &
  31.0\% &
  0.51 &
  0.50 &
  \underline{0.63} &
  \underline{0.63} &
   \underline{0.70} &
  \underline{0.70} &
  12.9 \\
Pocket2Mol &
  -5.14 &
  -4.70 &
  -6.42 &
  -5.82 &
  -7.15 &
  -6.79 &
  48.4\% &
  51.0\% &
  \textbf{0.56} &
  \textbf{0.57} &
 \textbf{ 0.74} &
 \textbf{ 0.75} &
  0.69 &
  \textbf{0.71} &
  24.3 \\
TargetDiff &
  -5.47 &
  \underline{-6.30} &
  -6.64 &
  -6.83 &
  -7.80 &
  -7.91 &
  58.1\% &
  59.1\% &
  0.48 &
  0.48 &
  0.58 &
  0.58 &
  \textbf{0.72} &
  \textbf{0.71} &
  20.5 \\
DecompDiff &
  -4.85 &
  -6.03 &
  -6.76 &
  \underline{-7.09} &
  \underline{-8.48} &
  \underline{-8.50} &
  \underline{64.8}\% &
  \textbf{78.6}\% &
  0.44 &
  0.41 &
  0.59 &
  0.59 &
  0.63 &
  0.62 &
  \underline{24.9} \\
\ourwork{} &
  \textbf{-7.02} &
  \textbf{-7.77} &
  \textbf{-7.95} &
  \textbf{-8.07} &
  \textbf{-8.59} &
  \textbf{-8.69} &
  \textbf{69.8}\% &
  \underline{76.4}\% &
  \underline{0.55} &
  \underline{0.56} &
  0.56 &
  0.56 &
  0.69 & %
  \underline{0.70} & %
  \textbf{27.7} \\
\bottomrule
\end{tabular}}
\end{table*}

\begin{table}[!htb]
\vspace{-0.2in}
\caption{\footnotesize Jensen-Shannon Divergence comparing bond distance distributions between reference molecules and generated molecules. Lower values indicate better performance. '-' represents single bonds, '=' represents double bonds, and ':' represents aromatic bonds. The first and second-place results are emphasized with bold and underlined
text, respectively.}
\label{tab:bonds_comparison}
\vskip 0.1in
\centering
\resizebox{\linewidth}{!}{
\begin{tabular}{@{}lcccccc@{}}
\toprule
\makecell{Bond} & \makecell{liGAN} & \makecell{AR} & \makecell{Pocket2} & \makecell{Target} & \makecell{Decomp} & \makecell{\ourwork{}} \\
& & & \makecell{Mol} & \makecell{Diff} & \makecell{Diff} & \\
\midrule
C-C  & 
    0.601 &
    0.609 &
    0.496 &
    \textbf{0.369}&
    \underline{0.371} & 
    0.384 \\
C=C  &
    0.665 &
    0.620 &
    0.561 &
    \underline{0.505} &
    0.539 &
    \textbf{0.501} \\
C-N  &
    0.634 &
    0.474 &
    0.416 &
    \underline{0.363} &
    \textbf{0.352} &
    0.365 \\
C=N  &
    0.749 &
    0.635 &
    0.629 &
    \textbf{0.550} &
    0.592 &
    \underline{0.559} \\
C-O  &
    0.656 &
    0.492 &
    0.454 &
    \underline{0.421} &
    \textbf{0.373} &
    0.422 \\
C=O  &
    0.661 &
    0.558 &
    0.516 &
    0.461 &
    \textbf{0.381} &
    \underline{0.430} \\
C:C  & 
    0.497 & 
    0.451 & 
    0.416 & 
    \underline{0.263} & 
    \textbf{0.258} & 
    0.269 \\
C:N  & 
    0.638 & 
    0.551 & 
    0.487 & 
    \textbf{0.235} &
    0.273 &
    \underline{0.252} \\
\bottomrule
\end{tabular}}
\end{table}

\begin{figure}
\vskip 0.2in
\centerline{\includegraphics[width=2.5in]{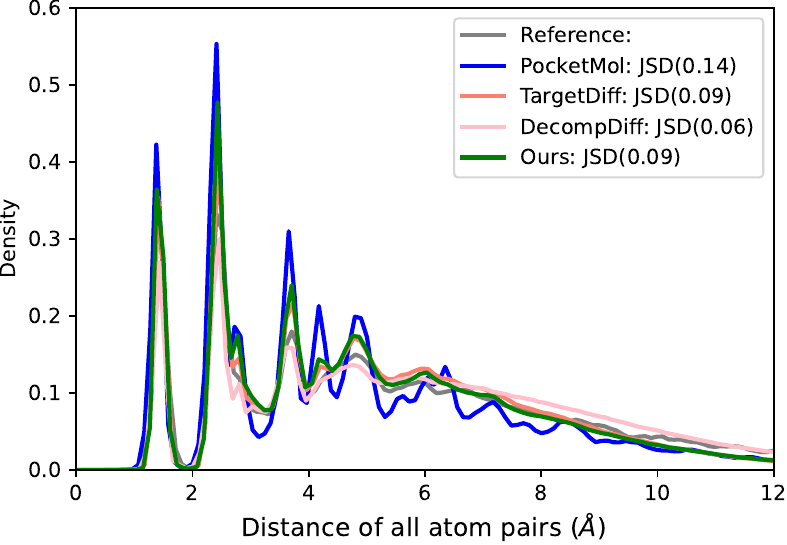}}
\caption{Comparison of distance distributions between all-atom distances of reference molecules in the test set (Reference) and distances in model-generated molecules. The Jensen-Shannon divergence (JSD) between these two distributions is calculated and reported.}
\label{fig:all_bonds}
\vskip -0.2in
\end{figure}
\vspace{-0.1in}
\subsection{Results}
\begin{table*}[!htp]
\centering
\caption{
Ablation analysis assessing the properties of generated molecules under different property guidance scenarios. The first and second-place results are emphasized with bold and underlined text, respectively.}
\label{tab:guidance_ablation}
\vskip 0.1in
\scalebox{1.0}{
\begin{tabular}{@{}l|cc|cc|cc|cc|cc|cc@{}}
\toprule
Methods &
  \multicolumn{2}{c|}{Vina Score ($\downarrow$)} &
  \multicolumn{2}{c|}{Vina Min ($\downarrow$)} &
  \multicolumn{2}{c|}{Vina Dock ($\downarrow$)} &
  \multicolumn{2}{c|}{QED ($\uparrow$)} &
  \multicolumn{2}{c|}{SA ($\uparrow$)} &
  \multicolumn{2}{c}{Hit ($\uparrow$)}
  \\
 &
  Avg. &
  Med. &
  Avg. &
  Med. &
  Avg. &
  Med. &
  Avg. &
  Med. &
  Avg. &
  Med. &
  Rate \\ \midrule
backbone &
  -5.47 &
  -6.30 &
  -6.64 &
  -6.83 &
  -7.80 &
  -7.91 &
  0.48 &
  0.48 &
  \underline{0.58} &
  \underline{0.58} &
  20.5 \\
backbone + BA Opt & \textbf{-7.35} & \textbf{-8.18} & \textbf{-8.38} & \textbf{-8.46} & \textbf{-9.04} & \textbf{-9.04} & 0.49 & 0.50 & 0.53 & 0.53 & 22.6 \\
backbone + QED Opt & -5.48 & -6.46 & -6.77 & -6.93 & -7.93 & -8.06 & \textbf{0.56} & \textbf{0.57} & \underline{0.58} & \underline{0.58} & \underline{24.5} \\
backbone + SA Opt & -5.22 & -6.03 & -6.40 & -6.57 & -7.53 & -7.73 & 0.47 & 0.48 & \textbf{0.61} & \textbf{0.60} & 19.4 \\
\ourwork{} &
\underline{-7.02} &
\underline{-7.77} &
\underline{-7.95} &
\underline{-8.07} &
\underline{-8.59} &
\underline{-8.69} &
\underline{0.55} &
\underline{0.56} &
0.56 &
0.56 &
\textbf{27.7}\\
\bottomrule
\end{tabular}
}
\end{table*}

\subsubsection{Performance of \ourwork{}}
\ourwork{} outperforms all baselines, including the reference molecules, in binding-related metrics and \textit{hit rate} (see Table \ref{tab:ablation_all}). This achievement is especially noteworthy when compared to the state-of-the-art DecompDiff model, which relies on external computation for informed priors in the denoising process. The success of our model, attained without requiring extra data, underscores the importance of effectively learning useful signals from the existing training set and skillfully guiding the diffusion denoising phase.

A key metric where \ourwork{} excels is the Vina Score, achieving a $22\%$ improvement over the state of the art (AR). This highlights its proficiency not only in generating molecules that bind effectively with proteins but also in producing high-affinity poses.
This is further substantiated by the fact that on average an impressive $69.8\%$ of the molecules generated by our model exhibit superior binding affinity compared to reference molecules, surpassing all other baselines. 
\looseness=-1

In terms of molecular properties, \ourwork{} shows remarkable performance, surpassing diffusion-based models in QED by $14.6\%$ and $22.2\%$ compared to TargetDiff and DecompDiff, respectively, while maintaining similar diversity levels. Although there's a slight decrease in SA, the metric remains within a reasonable range, indicating satisfactory synthesizability.
A detailed discussion of the challenges in optimizing SA is provided in Appendix \ref{sec:sa-challenges}.
Moreover, Appendix \ref{sec:statistical-significance} presents compelling evidence of statistically significant changes in guided properties upon the incorporation of guidance.
Furthermore, as shown in Table \ref{tab:bonds_comparison}, our method closely maintains the bond distribution in alignment with the backbone diffusion model, namely TargetDiff, surpassing the non-diffusion baselines. This outcome emphasizes the efficient synergy between our \textit{guide} model and the \textit{diffusion} model, markedly improving our ability to generate molecules with targeted properties while preserving molecular conformation. 

\begin{figure}
\centerline{\includegraphics[width=0.5\textwidth]{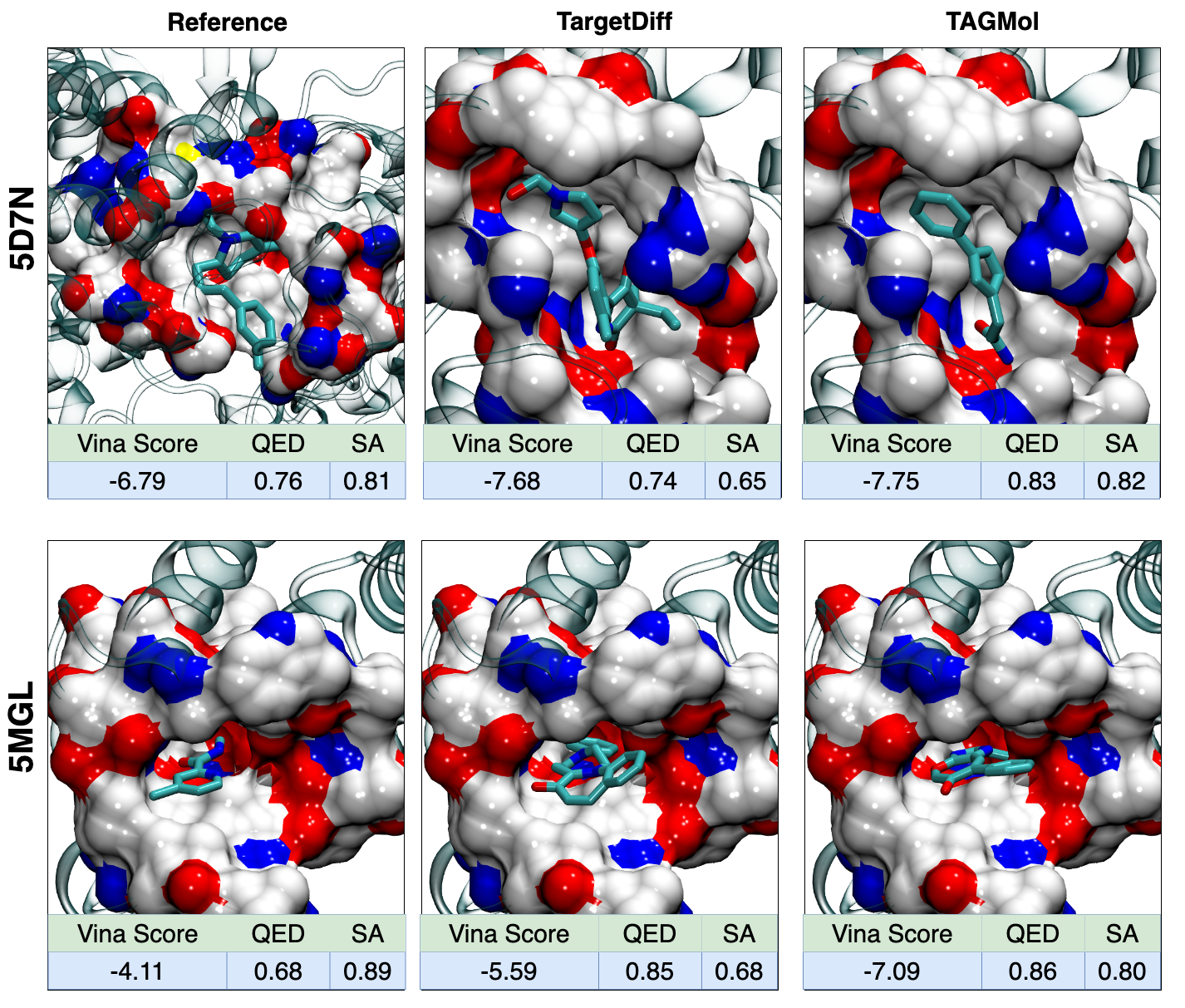}}
\vspace{-0.1in}
\caption{Visualization of reference molecules (left), alongside molecules generated by our backbone, TargetDiff (middle), and \ourwork{} (right), for two targets: 5D7N and 5MGL.}
\label{fig:ligands_viz}
\end{figure}

\ourwork{}, which focuses on optimizing multiple properties, achieves a state-of-the-art \textit{hit rate}, surpassing the backbone model—TargetDiff—by a considerable margin. This is evidenced by the higher fraction of molecules generated by our method that meet the \textit{hit criteria}. The collective results underscore the formidable capabilities of \ourwork{} in the SBDD domain. These advancements in critical areas of molecular generation and binding affinity demonstrate the model's potential to meet the intricate and diverse challenges of drug design. 
Figure \ref{fig:ligands_viz} presents examples of ligand molecules generated by our model, featuring valid structures and reasonable binding poses to the target.

\subsubsection{Single \& Multi Objective Guidance}
In Table \ref{tab:guidance_ablation}, we conduct an ablation study on our property guidance mechanism for multiple properties separately and all of them in tandem, demonstrating its effectiveness across various aspects. Specifically, we observe superior performance in each property when we provide respective guidance, showcasing the robustness of our guidance mechanism. We observe that when we provide guidance for one property, it does not result in a substantial deterioration in other properties. This highlights the collaborative mechanism between the \textit{denoising} model and the \textit{guide} model, emphasizing their combined power. 
Notably, a substantial enhancement is observed when we guide for binding affinity. This improvement is likely due to our approach of calculating binding affinity through docking software \cite{trott2010autodock}, which primarily depends on atomic distances.
Our guide model's architecture, with its inherent inductive bias in modeling atomic distances, effectively facilitates the explicit guidance for coordinate generation, contributing to this advancement.
Consequently, improvements in QED and SA scores are comparatively less pronounced, as these properties exhibit a 
relatively lower dependence on molecule's geometric configuration.
Lastly, when we provide guidance for all properties simultaneously, we witness substantial improvements across the majority of the properties, and achieves the highest \textit{hit rate}. However, a slight decrease in SA scores is observed, which, nonetheless, lies well within our \textit{hit criteria}. We also present additional visualization plots in Appendix~\ref{sec:multi_objective_rationale}.

\section{Conclusions}
In this work, we redefined the problem of drug generative modeling by extending our focus beyond merely optimizing binding activity. We emphasize the importance of optimizing additional properties and integrating these considerations into our problem formulation, even when training data is sparse. Secondly, we developed a optimized generative method, \ourwork{}, which effectively combines two key components: an \textit{SE(3) equivariant graph diffusion model} to accurately replicate the geometries found in the training set and a multi-objective guidance driven by \textit{SE(3) invariant GNN models}. This innovative combination enables our model to efficiently navigate the diffusion sampling space, focusing on properties of interest through the use of gradients. Lastly, our rigorous empirical evaluation demonstrates that \ourwork{} notably enhances metric performance, balancing the optimization of binding affinity with critical pharmacological properties such as QED and SA.

\textbf{Limitations and Future Work.} 
Although the present work shows promising results, we share some complementary directions worth investigating in the future.
First, improved representation of the system using additional inductive bias in terms of symmetry and coarse-graining could furnish more accurate predictions of different properties that are structure dependent.
Second, model compression and parameter pruning could speed-up inference without compromising performance.
Third, integrating constraints on discrete attributes into gradient-based optimization may enable the modeling of complementary properties, such as enforcing hydrogen bonds between specific residues within the pocket, a capability our current algorithm lacks. Fourth, exploring reduced noise sampling strategies or guidance mechanisms for generating more precise conformations could lead to reduced strain energy and the production of stable molecules.
Finally, it would be worth investigating physics-based techniques, such as molecular simulations, to allow generalization to larger family of systems.\looseness=-1

\section{Broader Impact}
Drug discovery is an area that can have broad impact on humanity. Accelerating the discovery of novel drugs can ensure more inclusive healthcare worldwide and can have impact especially on reducing the disparity within and across countries. In the development of our methodology, we have taken care to ensure its ethical application, fully acknowledging that there exists inadvertent potential for misuse. Specifically, there is a risk that our optimization algorithms could be repurposed for creating drugs with detrimental side effects. This underscores the importance of exercising caution and responsibility in the deployment of such algorithms in practical scenarios to prevent any unintended negative consequences from their use.
\bibliography{references}
\bibliographystyle{icml2024}
\newpage
\appendix
\onecolumn

\section{Notation Summary}
\label{sec:notations}
\begin{table}[htbp]
\caption{Notations}
\label{tab:notations}
\vskip 0.1in
\centering
\begin{tabular}{l|c|c}
\toprule
Notation & Explanation & Domain \\
\midrule
$N_{\gP}$ & Number of protein atoms & $\in \sR^1$ \\
$N_{f}$ & Feature dimension of protein atom & $\in \sR^1$ \\
$\rvx^{\gP}_{i}$ & $i$-th protein atom coordinate & $\in \sR^3$ \\
$\rvv^{\gP}_{i}$ & $i$-th protein atom feature & $\in \sR^{N_{f}}$ \\
$N_{\gM}$ & Number of ligand atoms & $\in \sR^1$ \\
$K$ & Number of ligand atom types & $\in \sR^1$ \\
$\rvx^{\gM}_{i}$ & $i$-th ligand atom coordinate & $\in \sR^3$ \\
$\rvx^{\gM}_{i,t}$ & $i$-th ligand atom coordinate at diffusion time step $t$ & $\in \sR^3$ \\
$\rvv^{\gM}_{i}$ & $i$-th ligand atom \textit{one-hot} representation & $\in \sR^K$ \\
$\rvv^{\gM}_{i,t}$ & $i$-th ligand atom \textit{one-hot} representation at diffusion time step $t$ & $\in \sR^K$ \\
$\rmX^{\gM}$ & Matrix representation of all ligand atom coordinates & $\in \sR^{N_{\gM}\times 3}$ \\ 
$\rmV^{\gM}$ & Matrix representation of all ligand atom types &  $\in \sR^{N_{\gM}\times K}$\\
$\rmX^{\gP}$ & Matrix representation of all protein atom coordinates & $\in \sR^{N_{\gP}\times 3}$ \\ 
$\rmV^{\gP}$ & Matrix representation of all protein atom features &  $\in \sR^{N_{\gP}\times N_f}$\\
$\beta_{t}$ & Variance Schedule for the diffusion model &  $\in \sR^{1}$ \\
$q$ & Diffusion noising transition function & - \\
$\theta$ & Parameters of generative model & - \\
$p_{\theta}$ & Denoising diffusion transition function &  - \\
$\sY$ & Set of properties to be optimized during generation &  - \\
$\rvy$ & A property $\rvy \in \sY$ &  - \\
$\phi_\rvy$ & Parameters of guide corresponding to $\rvy$ &  - \\
$\phi_\sY$ & Set of guide parameters corresponding to $\forall \rvy \in \sY$ &  - \\
$s_{\rvy}$ & Guide strength for property $\rvy$ &  $\in \sR^1$ \\
$s^{opt}_{\rvy}$ & Optimal Guide strength for property $\rvy$ &  $\in \sR^1$ \\
\bottomrule
\end{tabular}
\end{table}
\section{Related Work}
\label{sec:related_work}

\paragraph{Neural Drug Design}
Neural models have significantly changed the landscape for drug design that was previously dominated by computational techniques such as MD simulations and Docking \cite{Alonso2006CombiningDA}. 
It has gone through multiple paradigm shifts in the recent past - ranging from using only 1D representations (SMILES) \cite{Bjerrum2017MolecularGW, gomez2018automatic} to 2D representations (molecular graphs) \cite{Liu2018ConstrainedGV, Zhou2018OptimizationOM} to 3D representations \cite{skalic2019shape, Ragoza2020LearningAC} (molecular coordinates). 
Recent works have also showcased the importance of using target-aware models for practical drug design. 
Advancements in target-aware drug design, particularly in the realm of text and graph-based generative methods \cite{eckmann2022limo, stanton2022accelerating, chenthamarakshan2020cogmol, lee2023MOOD}, have made significant strides. 
However, these methods often generate ligands without considering the 3D structure of the target protein.
TargetDiff \cite{guan2023d} and DecompDiff \cite{guan2023decompdiff} are two diffusion-based models that consider the 3D structure of the target protein.
DecompDiff extends TargetDiff by decomposing the task into generating the arms and scaffold while explicitly considering bonds between the atoms during the diffusion process which is ignored by TargetDiff.
Fragment-based generation methods such as \citep{ghorbani2023autoregressive} allow for controlled generation, ensuring that only certain types of molecular fragments are present in the generated ligand. 
But they still fall behind the performance of diffusion-based approaches in terms of their binding affinity.

\paragraph{Molecular Property Prediction}
Predicting the properties of molecules through physio-chemical experiments is an expensive and time-consuming solution, that is unsuitable for getting intermediate feedback on AI-designed molecules. 
Using neural models has allowed high-quality prediction of various molecular properties in an automated fashion. 
Neural methods use different molecular representations to predict molecules, including SMILES representation with pre-trained transformers \cite{molxpt} or molecular structures \cite{Yang2019AnalyzingLM, zhou2023unimol} for predicting the properties of interest. 

\paragraph{Molecular Property Optimization}
Prior methods for property optimization have focused on using Reinforcement learning to generate molecules with desired properties \cite{you2018graph,zhou2019optimization,zhavoronkov2019deep, jeon2020autonomous,olivecrona2017molecular}.
However, RL methods are often computationally expensive and challenging to optimize due to the vast search space. 
LIMO \cite{eckmann2022limo} uses Variational Auto Encoders (VAEs) and gradient gudiance on MLP-based property predictors but does not consider the 3D structure of the target protein or the generated ligand. \cite{lee2023MOOD} introduced a novel graph-based diffusion model for out-of-distribution (OOD) generation, enhancing explorability through OOD-controlled reverse-time diffusion and property-guided sampling, albeit focusing solely on 2D molecular representations, unlike the target-aware 3D structure generation with multi-objective guidance featured in our work.

\section{Property Distribution in Training Data}
\begin{figure}[!htp]
\vskip 0.0in
\centerline{\includegraphics[width=\textwidth]{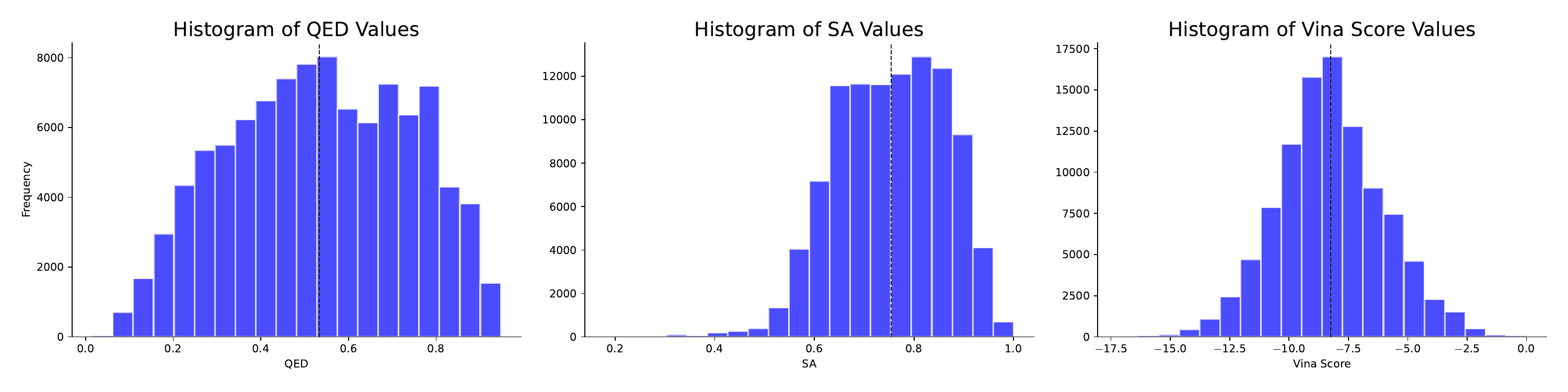}}
\caption{Distribution of QED, SA, and Vina Score properties in the training split, with the mean values represented by dotted lines}
\label{fig:property_train_density}
\vskip -0.2in
\end{figure}

Figure \ref{fig:property_train_density} illustrates the distribution of QED, SA, and Vina Scores within the training split. From the figure, it is evident that a substantial proportion of molecules exhibit low binding and fall below the hit criteria for QED. 

\section{Training and Evaluation of Guide}
\label{sec:train_eval_guide}
\subsection{Training}

The models are trained with a batch size of 4 to minimize the Root Mean Square Error (RMSE) loss, utilizing a gradient descent algorithm. The initial learning rates were set at 0.001 for the \textit{binding affinity} and \textit{QED guides}, and at 5e-4 for the \textit{SA guide}. Additionally, to maintain the stability of the training process, the gradient norms were clipped at a maximum value of 8. The learning rate underwent exponential decay by a factor of 0.6, with a lower bound set at 1e-5. This decay was triggered in the absence of any improvement in the validation loss across 10 consecutive evaluations. Evaluations were conducted every 2000 training steps, and convergence of all models was achieved within 200,000 steps. All experiments were conducted using the NVIDIA A100 GPU. 

\subsection{Evaluation}

The performance of each guide model was evaluated for their respective property predictions using the same test dataset.
As outlined in \S~\ref{sec:datasets}, we employed the same split to evaluate our \textit{binding affinity guide} model. While the train-test split ensures the uniqueness of the protein-ligand complexes between the sets, 34 ligands are common to both, leading to potential train-test leakage. Consequently, for \textit{QED} and \textit{SA} guides that depend solely on ligands, we have excluded these 34 overlapping ligands from the test split (No overlaps). Table \ref{tab:guide_eval} presents the performance evaluation of each \textit{guide} model, indicating high R$^2$ values on the original test dataset.
After removing the overlapping ligands from the test set, the performance metrics remained consistent, confirming the model's effective generalization to unseen ligands. While we report the results on fair test split, it's crucial to note that, diverging from the traditional machine learning training and testing process, we train the guide to direct the generated latent space of the diffusion model.

\begin{table}[!htp]
\centering
\caption{Evaluation of the guide model on the original test set with unique protein-ligand complexes, and on a modified test set (No Overlaps) where 32 overlapping ligands have been removed.}
\label{tab:guide_eval}
\vskip 0.15in
\begin{tabular}{@{}l|cc|cc@{}}
\toprule 
   \multirow{2}{*}{{\begin{tabular}[c]{@{}l@{}}Target \\ Property\end{tabular}}}
   & \multicolumn{2}{c|}{Test Set (Original)} & \multicolumn{2}{c}{Test Set (No Overlaps)} \\ 
    \cmidrule(l){2-3} \cmidrule(l){4-5}
    & RMSE        & R$^2$      & RMSE            & R$^2$          \\ \midrule
BA  & 0.642       & 0.953                     & 0.652           & 0.951                         \\
QED & 0.067       & 0.892                     & 0.079           & 0.850                          \\
SA  & 0.054       & 0.848                     & 0.065           & 0.821                         \\ \bottomrule
\end{tabular}
\end{table}

\section{Additonal Experiments}

\subsection{Search for optimal Guide Strength}
\label{sec:guide-strength-search}
To determine the optimal Guide Strength ($s$), a grid search was conducted over the set of values: $\{0, 1, 2, 5, 10, 20, 50\}$.
Here, $s=0$ represents the scenario without guidance, included for comparative purposes to gauge guide-induced enhancements.
The generation process was limited to 50 molecules for a limited set of 4 targets, to accommodate computational constraints.
Initially, for each property, we find the optimal guide strength $s^{opt}_\rvy$ over the grid. 
Once the optimal values for single objective are tuned, for multi-objective guidance, we obtained optimal $s$ values by recalibrating the $s^{opt}_\rvy$ with $w_{\rvy}$, as detailed in \S-\ref{sec:expt-setup}.

Table \ref{tab:combined_guidance_lr} presents the results of our tuning across different guide strengths ($s$) mentioned in the grid.
It is evident that non-zero $s$ values yield improved property metrics, underscoring the efficacy of guidance.
For Binding Affinity (BA) guidance, optimal $s$ values emerge as 1 and 2, based on average and median Vina Scores, respectively.
Considering both Vina Min and Vina Dock values, $s = 2$ is selected for superior outcomes.
In the case of QED guidance, an $s$ value of 20 distinctly outperforms others in terms of both average and median QED scores.
With SA guidance, although $s = 20$ and $s = 50$ appear to be optimal, the validity\cite{guan2023decompdiff} of generated molecules takes a low value of 67\% and 46\%, respectively.
Consequently, $s = 5$ is chosen, providing the second-best SA values, with 84\% of molecules being valid and only a slight decrease in average SA value by 0.02.
In summary, the selected Guide Strengths are: $s^{opt}_{ba} = 2$, $s^{opt}_{qed} = 20$, and $s^{opt}_{sa} = 5$.

\begin{table}[ht]
\centering
\caption{
Search for optimal Guide Strength ($s$), conducted individually for each Guide Model (single-objective setting).
To minimize computational demands, the generation is limited to 50 samples across 4 targets.
Only the metrics of the property being guided are included.
Symbols ($\uparrow$) / ($\downarrow$) denote preference for higher or lower values, respectively.
Top results are highlighted in bold for first place and underlined for second.
Italicized figures indicate instances where over 25\% of molecules generated under the specified guide strength were not valid.
}
\label{tab:combined_guidance_lr}
\vskip 0.15in
\begin{tabular}{@{}c|cc|cc|cc|cc|cc@{}}
\toprule
  &
  \multicolumn{6}{c|}{BA Guide} & \multicolumn{2}{c|}{QED Guide} & \multicolumn{2}{c}{SA Guide} \\ \cmidrule(l){2-7} \cmidrule(l){8-9} \cmidrule(l){10-11}
  $s$ & \multicolumn{2}{c|}{Vina Score ($\downarrow$)} &
  \multicolumn{2}{c|}{Vina Min ($\downarrow$)} &
  \multicolumn{2}{c|}{Vina Dock ($\downarrow$)} &
  \multicolumn{2}{c|}{QED ($\uparrow$)} &
  \multicolumn{2}{c}{SA ($\uparrow$)} \\
  & Avg. & Med. &  Avg. & Med. & Avg. & Med. &  Avg. & Med. & Avg. & Med.  \\ \midrule
0                     & -5.37                                 & -6.17                                   & -6.31                               & -6.83                                 & -7.65                                & -7.67                                  & 0.51                          & 0.54                            & 0.59                         & 0.59                           \\
1                     & \textbf{-7.82}                        & \underline{-7.93}                             & \underline{-8.16}                         & \underline{-8.24}                           & -8.62                                & -8.64                                  & 0.52                          & 0.53                            & 0.59                         & 0.59                           \\
2                     & \underline{-7.69}                           & \textbf{-8.08}                          & \textbf{-8.33}                      & \textbf{-8.40}                        & \textbf{-8.86}                       & \textbf{-8.89}                         & 0.52                          & 0.53                            & 0.60                         & 0.59                           \\
5                     & -7.55                                 & -7.74                                   & -8.02                               & -8.04                                 & \underline{-8.67}                          & \underline{-8.65}                            & \underline{0.55}                    & \underline{0.58}                      & \underline{0.62}                   & \underline{0.62}                     \\
10                    & -7.11                                 & -7.09                                   & -7.53                               & -7.33                                 & -8.42                                & -8.34                                  & 0.54                          & 0.54                            & 0.60                         & 0.59                           \\
20                    & -6.85                                 & -6.69                                   & -7.03                               & -6.87                                 & -8.16                                & -8.14                                  & \textbf{0.58}                 & \textbf{0.60}                   & \textit{\textbf{0.64}}                & \textit{\textbf{0.63}}                  \\
50                    & -6.11                                 & -5.96                                   & -5.68                               & -5.91                                 & -7.73                                & -7.44                                  & \underline{0.55}                    & 0.55                            & \textit{\textbf{0.64}}                & \textit{\textbf{0.63}}                  \\ \bottomrule
\end{tabular}
\end{table}

\subsection{Ablation on Single and Multi-Objective Guidance}
\label{sec:single-dual-multi-ablation}
To assess the impact of guidance on various combinations of properties, we conducted a detailed ablation study. Initially, we present results for single-objective guidance on different properties, then proceed to guidance for two properties at a time, and ultimately, guidance incorporating all properties. 
Our results, highlighted in Table \ref{tab:guidance_ablation_all}, indicate that single-objective guidance maximizes improvement for the targeted property. When applying guidance to two properties, the enhancements for each are moderate, illustrating a trade-off compared to separate guidance. This trade-off persists with guidance across all properties. However, given the multifaceted nature of drug discovery, where the goal is to produce molecules with high binding affinity and desired properties like QED and SA, our approach significantly enhances \textit{hit rate} and most properties over no guidance. This underscores the effectiveness of guidance in navigating regions that are crucial in discovery.

\begin{table}[!htb]
\centering
\caption{
Extensive ablation analysis assessing the properties of generated molecules under different property guidance scenarios. The first and second-place results are emphasized with bold and underlined text, respectively.}
\label{tab:guidance_ablation_all}
\vskip 0.15in
\begin{tabular}{ccc|cc|cc|cc|cc|cc|cc}
\toprule
 $w_{ba}$ & $w_{qed}$ & $w_{sa}$ &
  \multicolumn{2}{c|}{Vina Score ($\downarrow$)} &
  \multicolumn{2}{c|}{Vina Min ($\downarrow$)} &
  \multicolumn{2}{c|}{Vina Dock ($\downarrow$)} &
  \multicolumn{2}{c|}{QED ($\uparrow$)} &
  \multicolumn{2}{c|}{SA ($\uparrow$)} &
  \multicolumn{2}{c}{Hit ($\uparrow$)}
  \\
 & & &
  Avg. &
  Med. &
  Avg. &
  Med. &
  Avg. &
  Med. &
  Avg. &
  Med. &
  Avg. &
  Med. &
  Rate \\ \midrule
0  & 0 & 0 & -5.47 & -6.30 & -6.64 & -6.83 & -7.80 & -7.91 & 0.48 & 0.48 & 0.58 & 0.58 &  20.5\\
\midrule
1  & 0 & 0 & \textbf{-7.35} & \textbf{-8.18} & \textbf{-8.38} & \textbf{-8.46} & \textbf{-9.04} & \textbf{-9.04} & 0.49 & 0.5 & 0.53 & 0.53 & 22.6 \\
0  & 1 & 0 & -5.48 & -6.46 & -6.77 & -6.93 & -7.93 & -8.06 & \textbf{0.56} &\textbf{0.57} & 0.58 & 0.58 & 24.5 \\
0  & 0 & 1 & -5.22 & -6.03 & -6.4 & -6.57 & -7.53 & -7.73 & 0.47 & 0.48 & \textbf{0.61} & \textbf{0.6} & 19.4\\
\midrule
0.5  & 0.5 & 0 & \underline{-7.11} & \underline{-7.96} & \underline{-8.13} & -8.21 & \underline{-8.82} & \underline{-8.85} & \textbf{0.56} & \textbf{0.57} & 0.55 & 0.55 & \underline{26.1} \\
0.5  & 0 & 0.5 & -7.2 & -7.95 & -8.16 & \underline{-8.26} & -8.83 & -8.84 & 0.5 & 0.51 & 0.55 & 0.55 & 24.7 \\
0  & 0.5 & 0.5 & -5.43 & -6.34 & -6.63 & -6.86 & -7.85 & -7.97 & \underline{0.55} & \underline{0.56} & \underline{0.59} & \underline{0.59} & 24.1 \\
\midrule
0.34  & 0.33 & 0.33 & -7.02 & -7.77 & -7.95 & -8.07 & -8.59 & -8.69 & \underline{0.55} & \underline{0.56} & 0.56 & 0.56 & \textbf{27.7}\\
\bottomrule
\end{tabular}%
\end{table}

\subsection{Multi-Objective Rationale}
\label{sec:multi_objective_rationale}

\begin{figure*}[!htp]
\vskip 0.2in
\begin{center}
\centerline{\includegraphics[width=\textwidth]{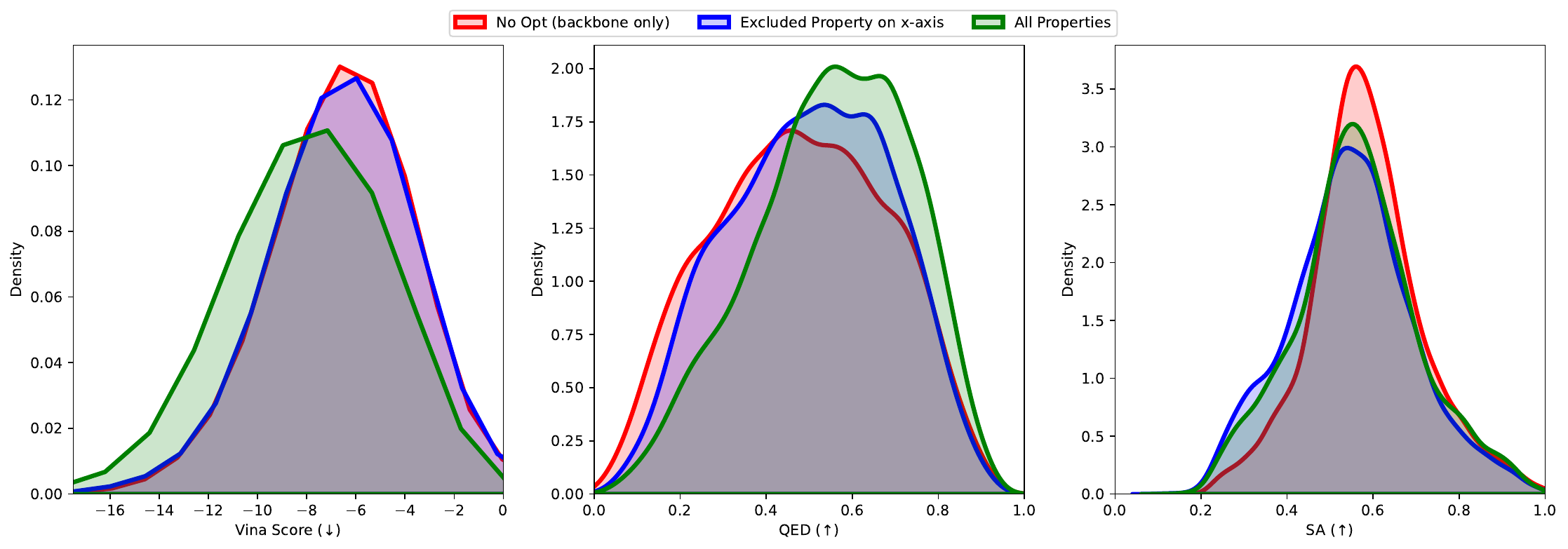}}
\caption{Distribution of molecular properties in molecules generated by the backbone model without any guidance (No Opt), when guided by two properties while excluding the one indicated on the x-axis (Excluded Property on x-axis), and when guided by all three properties (All Properties). ($\downarrow$) denote properties where lower values are preferred, while ($\uparrow$) indicate properties where higher values are desirable.}
\label{fig:multi_opt_density}
\end{center}
\vskip -0.2in
\end{figure*}

Figure \ref{fig:multi_opt_density} presents the distribution of molecular properties for 10,000 generated molecules (100 per target, across 100 targets).
These molecules are generated using three methods: (i) the backbone model without any guidance; (ii) guidance using only two properties, leaving out the third one indicated on the x-axis; and (iii) guidance using all three properties.
Optimizing all the properties helps improve the Vina Score and QED compared to the `No Opt' baseline. 
Although it doesn't help improve the SA, it is still helpful, as excluding the SA property in optimization worsens the SA scores.
Thus, it is useful to provide guidance for all three properties.

\subsection{Challenges in SA optimization}
\label{sec:sa-challenges}
The SA score serves as a comprehensive metric for assessing synthetic feasibility, considering various non-standard structural features such as large rings, unconventional ring fusions, stereocomplexity, and overall molecule size \cite{ertl2009estimation}. However, diffusion-based generative models face a limitation in accurately positioning atoms, leading to unrealistic molecular topologies, especially the formation of large rings, which can negatively impact SA scores \cite{peng2023moldiff}.

As evident from Table \ref{tab:rings}, the diffusion-based models exhibit a subpar distribution of ring sizes, notably characterized by a prevalence of larger rings. In \ourwork{}, the provision of multiple guidance signals to coordinates increases the likelihood of forming large rings. Consequently, this propensity towards larger ring formations contributes to either a drop in SA values or minimal improvement despite the guidance provided.

\begin{table}[htbp]
\caption{Distribution of ring sizes in reference and generated molecules, expressed as percentages}
\label{tab:rings}
\vskip 0.15in
\centering
\begin{small}
\begin{tabular}{lccccccc}
\toprule
\makecell{Ring} & \makecell{Ref} & \makecell{Pocket2} & \makecell{Target} & \makecell{Decomp} & \makecell{\ourwork{}} \\
 Size & & \makecell{Mol} & \makecell{Diff} & \makecell{Diff} & \\
\midrule
3 & 1.7\% & 0.1\% & 0.0\% & 2.9\% & 0.0\% \\
4 & 0.0\% & 0.0\% & 2.8\% & 3.7\% & 2.4\% \\
5 & 30.2\% & 16.4\% & 30.8\% & 30.8\% & 26.9\% \\
6 & 67.4\% & 80.4\% & 50.7\% & 45.6\% & 48.6\% \\
7 & 0.7\% &  2.6\% & 12.1\% & 11.6\% & 15.0\% \\
8 & 0.0\% & 0.3\% & 2.7\% & 2.3\% & 3.7\% \\
$\geq$9 & 0.0\% & 0.2\% & 0.9\% & 3.1\% & 3.4\% \\
\bottomrule
\end{tabular}
\end{small}
\end{table}

\subsection{Statistical Significance of Guidance}
\label{sec:statistical-significance}

To demonstrate that the integration of guidance contributes to statistically significant changes in the results, we employ Paired $t$-test (for QED, SA, Vina Score, Vina Min and Vina Dock) and Chi-square test (for \textit{hit rate}), comparing our model, \ourwork{}, against the backbone model, TargetDiff.
For the Paired $t$-test, we group the generated samples by target protein and compute the average scores for each property, yielding 100 pairs for comparison in our case.
Our null hypothesis posits that there is no difference in the average values of a given property produced by the two models across various protein targets.
The $p$-values for the guided properties QED, SA, and Vina Score were remarkably low, at 4.65E-33, 5.00E-13, and 7.05E-10, respectively.
Similarly, for Vina Min and Vina Dock, the p-values were 2.89E-15 and 4.38E-06, respectively.

On the other hand, for the \textit{hit rate}, we utilize the Chi-square test, since the $t$-test may not be the suitable test.
Here, outcomes for a protein target are categorized as either ``a greater number of molecules generated by \ourwork{} satisfied the \textit{hit criteria} compared to those by TargetDiff'' or the reverse, with the null hypothesis being an equal likelihood of occurrence, implying no difference between the models.
The obtained $p$-value of 5.73E-07 further supports our findings.

Given that the $p$-values across all tests are substantially lower the than the conventional threshold of 0.05, we can confidently reject the null hypothesis, affirming that the guidance incorporation leads to highly statistically significant differences in the outcomes.

\subsection{Benchmarking with PoseCheck}
\label{app:posecheck}
Expanding our evaluation, we utilized PoseCheck \cite{harris2023posecheck} to evaluate the 3D poses generated by the models. Our evaluation is focused on steric clashes and strain energy. Steric clashes quantify instances where the pairwise distance between a protein and ligand atom is below the sum of their van der Waals radii, with a clash tolerance of 0.5 Å. Meanwhile, strain energy represents the internal energy accumulated within a ligand due to conformational changes upon binding.

In Table \ref{tab:posecheck_gen} and \ref{tab:posecheck_redock}, we present the mean values of steric clashes and the median values of strain energy for the generated molecules both before and after docking. In our discussion, we prioritize median values for strain energy as they offer greater representativeness in this context, especially given the presence of significant outliers.

\ourwork{} demonstrates superior performance compared to diffusion-based models TargetDiff and DecompDiff in terms of clashes, highlighting the efficacy of our binding affinity guide.
However, \ourwork{} exhibits shortcomings concerning strain energy. As elucidated in PoseCheck, diffusion-based methodologies often yield elevated strain energies, a trend also observed in TAGMol due to minor errors in atom placement resulting from coordinate guidance. We hypothesize that integrating strain energy guidance during the generation phase could yield more stable molecules. We plan to investigate this in our future work.

\begin{table}[!htp]
\caption{Evaluation of poses of generated molecules on biophysical benchmarks, PoseCheck. ($\uparrow$) / ($\downarrow$) indicates whether a larger or smaller value is preferable. The first-place results are emphasized in bold.}
\label{tab:posecheck_gen}
\vskip 0.15in
\centering
\begin{tabular}{l|c|c}
\toprule
Methods  & Clashes ($\downarrow$) & Strain Energy ($\downarrow$) \\ \midrule
TargetDiff & 9.19        & 1258.02           \\
DecompDiff & 12.53       & \textbf{539.89}            \\
\ourwork{}     & \textbf{6.05}        & 2143.03           \\
\bottomrule
\end{tabular}
\end{table}

\begin{table}[!htp]
\caption{Evaluation of poses of generated molecules on biophysical benchmarks, PoseCheck, after redocking. ($\uparrow$) / ($\downarrow$) indicates whether a larger or smaller value is preferable. The first-place results are emphasized in bold.}
\label{tab:posecheck_redock}
\vskip 0.15in
\centering
\begin{tabular}{l|c|c}
\toprule
Methods  & Clashes ($\downarrow$) & Strain Energy ($\downarrow$) \\ \midrule
Reference  & 3.5         & 118.17            \\ 
\midrule
TargetDiff & 5.9         & 602.00            \\ 
DecompDiff & 5.75        & \textbf{441.61}            \\ 
\ourwork{}     & \textbf{5.36}        & 709.35            \\
\bottomrule
\end{tabular}
\end{table}

\subsection{Additional baselines}

Enhancing our evaluation, we have incorporated two new methods, IPDiff\footnote{Since these are contemporary works that will appear in ICLR 24, a comprehensive comparison could not be conducted.} \cite{huang2024proteinligand} and DecompOPT\footnotemark[\value{footnote}] \cite{zhou2024decompopt}. These methods, which have been recently proposed, introduce additional supervised signals to enhance the generative process.

IPDiff first pre-trains a Graph Neural Network (GNN) using binding affinity signals as supervision. It then leverages this pretrained network to integrate interactions between the target protein and the molecular ligand into the forward process.

On the other hand, DecompOPT adopts an optimization approach based on a controllable and decomposed diffusion model. Initially, it generates a set of molecules using DecompDiff, which are subsequently broken down for additional local optimization via evolutionary algorithms. Throughout the evolutionary process, supervised signals such as binding affinity, QED, and SA for decomposed substructures are recalculated to retain the most effective substructures.

By including these new methods in our evaluation, we aim to provide a more comprehensive comparison that captures the nuances of utilizing additional supervised signals, particularly binding affinity, QED and SA, in molecular property optimization. We believe that this expanded evaluation enhances the robustness and depth of our study.

Table \ref{tab:ablation_add_baselines} presents a comparative analysis of the current state-of-the-art diffusion-based models, including two novel methods described earlier. \ourwork{} showcases superior performance across Vina Score, Vina Min, and QED, while maintaining comparable diversity, highlighting the effectiveness of our guidance. \ourwork{}'s significant outperformance of DecompOPT across various metrics is particularly notable, especially considering that DecompOPT refines molecules generated by DecompDiff. Moreover, \ourwork{} surpasses IPDiff in all binding affinity-related metrics, despite IPDiff incorporating binding affinity signals during training. In summary, \ourwork{}'s success stands out when compared to models employing additional supervised signals, highlighting the effectiveness of our guidance scheme.
\begin{table*}[!htb]
\caption{\footnotesize Comparison of various properties between reference molecules and those generated by our model and other new baselines. ($\uparrow$) / ($\downarrow$) indicates whether a larger or smaller number is preferable. The first and second-place results are emphasized with bold and underlined text, respectively.}
\label{tab:ablation_add_baselines}
\vskip 0.1in
\resizebox{\linewidth}{!}{
\begin{tabular}{@{}l|cc|cc|cc|cc|cc|cc|cc@{}}
\toprule
Methods &
  \multicolumn{2}{c|}{Vina Score ($\downarrow$)} &
  \multicolumn{2}{c|}{Vina Min ($\downarrow$)} &
  \multicolumn{2}{c|}{Vina Dock ($\downarrow$)} &
  \multicolumn{2}{c|}{High Affinity ($\uparrow$)} &
  \multicolumn{2}{c|}{QED ($\uparrow$)} &
  \multicolumn{2}{c|}{SA ($\uparrow$)} &
  \multicolumn{2}{c}{Diversity ($\uparrow$)} \\
 &
  Avg. &
  Med. &
  Avg. &
  Med. &
  Avg. &
  Med. &
  Avg. &
  Med. &
  Avg. &
  Med. &
  Avg. &
  Med. &
  Avg. &
  Med. 
  \\ \midrule
Reference &
  -6.36 &
  -6.46 &
  -6.71 &
  -6.49 &
  -7.45 &
  -7.26 &
  - &
  - &
  0.48 &
  0.47 &
  0.73 &
  0.74 &
  - &
  -  \\ \midrule
TargetDiff &
  -5.47 &
  -6.30 &
  -6.64 &
  -6.83 &
  -7.80 &
  -7.91 &
  58.1\% &
  59.1\% &
  0.48 &
  0.48 &
  0.58 &
  0.58 &
  \underline{0.72} &
  \underline{0.71} \\
DecompDiff &
  -4.85 &
  -6.03 &
  -6.76 &
  -7.09 &
  -8.48 &
  -8.50 &
  64.8\% &
  78.6\% &
  0.44 &
  0.41 &
  \underline{0.59} &
  \underline{0.59} &
  0.63 &
  0.62 \\
IPDiff &
  \underline{-6.42} &
  \underline{-7.01} &
  \underline{-7.45} &
  -7.48 &
  -8.57 &
  -8.51 &
  69.5\% &
  75.5\% &
  0.52 &
  0.52 &
  \underline{0.59} &
  \underline{0.59} &
  \textbf{0.74} &
  \textbf{0.73} \\
DecompOPT &
  -5.87 &
  -6.81 &
  -7.35 &
  \underline{-7.78} &
  \textbf{-8.98} &
  \textbf{-9.01} &
  \textbf{73.5}\% &
  \textbf{93.3}\% &
  0.48 &
  0.45 &
  \textbf{0.65} &
  \textbf{0.65} &
  0.60 &
  0.60 \\
\ourwork{} &
  \textbf{-7.02} &
  \textbf{-7.77} &
  \textbf{-7.95} &
  \textbf{-8.07} &
  \underline{-8.59} &
  \underline{-8.69} &
  \underline{69.8}\% &
  \underline{76.4}\% &
  \textbf{0.55} &
  \textbf{0.56} &
  0.56 &
  0.56 &
  0.68 &
  0.69 \\
\bottomrule
\end{tabular}}
\end{table*}

\section{Time Complexity}
When executed on an NVIDIA A100 GPU, we observed a notable processing time trend for inference. Guidance for properties like QED and SA led to significant improvements (refer to Table \ref{tab:guidance_ablation_all}) with minimal time increases of 1.06x and 1.07x, respectively, compared to the backbone. However, guiding for Binding Affinity (BA) led to a more substantial time increase due to the inclusion of protein atoms, demanding extra computational effort compared to QED or SA guidance, which involves only ligand atoms. Ultimately, \ourwork{} recorded its longest processing time of 1.87x  when applying guidance for all properties. It is important to note that \ourwork{}  is still comparable to 1.72x of DecompDiff \cite{guan2023d} sampling time. 
\begin{table}[!htp]
\centering
\caption{
Inference time for different models
}
\label{tab:time_complexity}
\vskip 0.15in
\scalebox{1.0}{
\begin{tabular}{l|cccc}
\toprule
Method & Time (s)
  \\ \midrule
backbone & 938 & \\
backbone + BA opt & 1646 \\
backbone + QED opt & 995 \\
backbone + SA opt & 1007 \\
\ourwork{} & 1755 \\
\bottomrule
\end{tabular}}
\end{table}

\section{Pseudo Code for \ourwork{}}

This section provides a summary of the overall sampling procedures employing multiple guidances.

\begin{algorithm}[!htbp]
\caption{Psuedo Code of \ourwork{}}
\label{alg:sampling}
\begin{algorithmic}
\REQUIRE The protein binding site $\rmP$, generative model $\theta$, Property Predictors $\phi_{\sY}$.
\ENSURE Generate ligand molecule $\rmM$ that binds to the protein pocket \& optimized for properties $\sY$ 

\STATE Sample $N_{\gM}$ atoms based on a prior distribution relative to the pocket size.
\STATE Move CoM of protein atoms to zero.
\STATE Sample coordinates $\rmX^{\gM}_T$ and atom types $\rmV^{\gM}_T$ based on prior: 

$\rmX^{\gM}_T \sim \mathcal{N}(0, \mathbf{I})$

$\rmV^{\gM}_T = \texttt{one\_hot}(\argmax_i g_i)$, where $g \sim \text{Gumbel}(0, 1)$

\FOR {$t = T, T-1, \dots, 1$}
    \STATE Calculate $[\widehat{\rmX}^{\gM}_{0|t}, \widehat{\rmV}^{\gM}_{0|t}]$ using $\theta([\rmX^{\gM}_{t}, \rmV^{\gM}_{t}], t, \mathcal{P})$
    \STATE Calculate $\Tilde{\mu}_t\left(\rmX^{\gM}_{t}, \hat{\rmX}^{\gM}_{0|t}\right)$ according to posterior in Equation \ref{eq:posterior}
    \STATE Calculate $\delta$ according to Equation \ref{eq:multi_guide_term}
    \STATE Sample $\rmM_{t-1}$ according to the Equation \ref{eq:multi_guide_samp}
\ENDFOR
\end{algorithmic}
\end{algorithm}

\end{document}